\documentclass[proof]{WileyASNA-v1}

\articletype{Article Type}%

\received{}
\revised{}
\accepted{}

\begin{document}

\title{Heating of Millisecond Pulsars by Magnetic Field Decay\protect\thanks{urme.geppert@gmail.com.}}

\author[1]{U. Geppert*}

\authormark{U. Geppert}

\address{Könneritzstraße 1, D-04229 Leipzig \country{Germany}}

\corres{*U. Geppert, \email{urme.geppert@gmail.com}}

\abstract{Millisecond pulsars (MSPs) are believed to be very old neutron stars (NSs) whose age may exceed 
significantly $10^8$ yrs. Although cooling scenarios of isolated NSs predict for that age a surface temperature $T_s\sim 10^4$ K, 
observations of the nearest MSP J0437-4715 indicate $T_s$ well above that value.
Besides the heating of the polar cap surface by backflowing charged particles, Joule heating in the crust 
can contribute to the overall heat budget of MSPs.\\
Since the dipolar field component, derived from $P$ and $\dot{P}$ measurements, is much too weak for remarkable heating,
smaller-scale structures should be analysed whether they can supply the demanded heat.
For this purpose we study the small scale field structure of radio pulsars. 
Magnetic field components, significantly stronger than the dipolar one, may exist especially at the surface of MSPs.
We assign upper limits to the strength of single field components up to a multipolarity of $l=10$ and the corresponding 
deviations from axial symmetry $m \le l$. Arguments are provided that the decay of
the small-scale components with $l=3$ or $l=4$ of the crustal magnetic field may cause the relatively high surface temperature of isolated
MSPs.}

\maketitle

\footnotetext{\textbf{Abbreviations:} NS, neutron star; MSP, millisecond pulsar}

\section{Introduction}\label{sec1}
Millisecond pulsars (MSPs) are characterized by relatively small dipolar magnetic surface 
field strength and a corresponding small deceleration of their rotational 
velocity $\dot{P} \lesssim 10^{-19}$. This returns a high characteristic age $\tau_c=P/2\dot{P}$ (rotational period $P$
and its time derivative $\dot{P}$). 
Since MSPs frequently reside in globular clusters it is assumed that they are old objects with $\tau_{age}$ 
exceeding $\sim 10^8$ yr. They received their rapid rotation with rotational periods $P \lesssim 10$ ms during 
its life in a close binary system  by heavy accretion driven angular momentum transfer from a companion star.
Their characteristic age, however can be considered only as an upper limit to the true age \citep{M17}. 
Because the spin up time period in a Low Mass X-Ray Binary system (LMXB) complicates significantly the rotational history of the NS, 
sophisticated methods had to be used to determine their true age. \cite{KT10} developed such tools and 
concluded that the characteristic age can exceed (or fall below) the true age by a factor of about $10$.  \\
\noindent One of the best studied MSP is J0437 - 4714 detected by \cite{BTBB93}. Its surface temperature was subject
to many very detailed observations and analyses (see \cite{YNH20} and references therein). All of them report 
a $T_s\approx 5.1\ldots 5.5\times 10^5$K. \cite{B13} found $T_s$ exceeding $10^6$ K, however only in the small area
around the pole.This observation can certainly be attributed to the bombardment of the polar cap
with ultra-relativistic charged particles.\\
Different heating mechanism that
may responsible for a surface temperature $T_s \gtrsim 10^5$ K have been discussed by \cite{GR10}. These are (besides
some more exotic ones) frictional motion of superfluid vortices, rotochemical heating, magnetic field decay, 
and crust cracking. \cite{GR10} excluded magnetic field decay because the available magnetic energy is too small for 
the necessary heating.\\
\noindent Here we will discuss that under certain conditions higher multipole components of the (sub-) surface field
can well contribute a significant amount of heat, thereby providing an explanation for the $T_s$-observations of isolated MSPs.
 
\section{Upper limits of the magnetic field components $B_{\MakeLowercase{lm}}$}\label{sec2}
More than thirty years ago, J. Krolik published a formula that relates the strengths of the
pulsar surface magnetic 
field components to the spin-down luminosity $L= -4\pi^2I\dot{P}/P^3$, with $I$ being the pulsar moment of inertia.
The assignment of the rotational energy losses to the luminosity of magneto-multipole radiation provides the
following equation \citep{K91}:

\begin{equation}
\left<|B_{lm}|\right>\le \frac{S_{lm}}{m^{l-1/2}}\frac{(2l)!}{2^ll!}\left(\frac{Ic^{2l+1}}{2^{2l-3}\pi^{2l-1}}\right)^{1/2}
R_{NS}^{-(l+2)}\left(\dot{P}P^{2l-1}\right)^{1/2}\,
\label{eq:Krolik_1}
\end{equation}

\noindent where $\left<|B_{lm}|\right>$ designates the over solid angle averaged surface field components, 
$I$ the NS moment of inertia (set to $10^{45}$ g cm$^2$), and $R_{NS}$ the NS radius (set to $10^6$ cm). An
equal sign in Eq.~\ref{eq:Krolik_1} applies if the corresponding multipole component is responsible alone and 
completly for the spin down of the NSs.\\
\noindent Inserting the values of J0437-4715 $P=0.00576$ s, $\dot{P}=5.73\times 10^{-20}$ and $R_{NS}=13.5$ km,
and the $S_{lm}$-values up to $l,m=4$ as given by \cite{K91}, we
find the upper limits for the field components as shown in Tab.~\ref{tab:Blm}.

\begin{table*}
    \caption{Magnetic field components $B_{lm}$ (in Gauss) according to Eq. ~\ref{eq:Krolik_1} as functions of $l$ and $m$. 
     Note that only components with $m\neq 0$ contribute and the magnetic field starts with $l=1$.}
    \label{tab:Blm}
    \begin{center}
   \begin{tabular}{llcccccccc}
        \hline
        \hline
        $m$    &   $1$ & $2$ & $3$ & $4$ & $5$ & $6$ & $10$ \\
     $l$ &   &  &   \\
        \hline
 & & & & & & & &\\
1  &  $9.7\times 10^8$  \\
2  &  $1.4\times 10^{11}$   & $1.0\times 10^{11}$   \\
3  &  $2.4\times 10^{13}$ & $1.4\times 10^{13}$ & $1.2\times 10^{13}$ \\
4  &  $4.9\times 10^{15}$ & $2.0\times 10^{15}$ & $1.8\times 10^{15}$ & $1.7\times 10^{15}$ \\ 
5  &  $3.2\times 10^{17}$ & $3.2\times 10^{17}$ & $3.2\times 10^{17}$ & $3.2\times 10^{17}$ & $3.2\times 10^{17}$ \\
6  &  $7.1\times 10^{19}$ & $7.1\times 10^{19}$ & $7.1\times 10^{19}$ & $7.1\times 10^{19}$ & $7.1\times 10^{19}$ & $7.1\times 10^{19}$\\
10 &  $8.2\times 10^{36}$ & $8.2\times 10^{36}$ & $8.2\times 10^{36}$ & $8.2\times 10^{36}$ & $8.2\times 10^{36}$ & $8.2\times 10^{36}$ & $8.2\times 10^{36}$\\
 & & & & & & & &\\
        \hline
  \end{tabular}
  \end{center}
\label{tab:Blm}
\end{table*}

\noindent Note that components with $l >4 $ do not change significantly if $m$ increases from $m=1$ to $m\le l$.
Therefore, the strength of the single multipole field components is determined mainly by their variation in
meridional direction.\\
Of course, the large magnetic field components for $l > 4$ reflect only the shortness of the ``lever arm" of these 
multipoles, so that unrealistically strong magnetic fields would be necessary to realise the observed deceleration 
of the rotation. These values have no relevance for real physics in MSPs.  However, magnetic field components 
with $l < 5$ can play a role in the heating of NSs.

\section{Can the presence of components with $\MakeLowercase{l}\ge 3$ explain $T_s\gtrsim 10^5$ K?}\label{sec3}
For isolated NSs older than $10^6$ yrs the cooling process is dominated by photon emission from the surface \citep{PGW06}.
The surface temperature at an age $t > 10^9$ yrs is in the order of $10^4$ K. 
\cite{GR10} investigated whether magnetic field decay could be a possible heat source in MSPs. 
They compared the luminosity for $T_s=10^5$ K with the loss of magnetic energy due to field decay, assuming 
that this is the only heat source.
In this way, they gained a relationship between the averaged magnetic field $B_{rms}$ and 
the age of the MSP $t$ (in $10^7$ yrs)  $B_{rms}=10^{13}\sqrt{t_7}$ G. Thus, an almost $10^{10}$ yrs old
MSP requires an averaged magnetic field of $\approx 3\times 10^{14}$ G in order to have a detectable $T_s$.
Field components with $l>3$ could provide these strength, even if they participate only with one percent
or less in the total magnetic field configuration. In case of J0437-4715 the appropriate component could be the $l=4$ one.

\section{May small scale components $\MakeLowercase{l,m}\lesssim 4$ survive for more than $10^8$ years?}\label{sec5}
In a normal isolated NS the dipolar component of the surface magnetic field $B_s$ decays 
from a strength at birth of $\gtrsim 10^{13}$ G by about $4\ldots 5$ orders of magnitude 
up to an age of $10^{9}$ years \citep{GRPR23}.
The decay is driven by Ohmic decay and Hall drift in the crust and/or processes in the core as
e.g. ambipolar diffusion \citep{VGGPDG21,GDGI22}.\\
\noindent The larger the magnetic field gradients, the more rapid proceeds the dissipation. When the field
penetrates the whole NS this gradient is $\propto l/R_{NS}$. However, if the currents maintaining the (sub-) surface 
field circulate in the crust, the field gradient is set by the crust thickness and comparable at least for
all multipoles $l < 10$ \citep{SC87,K91}.\\
\noindent It is widely believed that isolated MSPs like J0437-4715 are spun up in a long lasting ($\sim 10^7$ yrs)
Roche lobe overflow phase, accreting matter from a companion star with accretion rates of about 
$10^{-9\ldots -10}$ M$_{\odot}/$ yr \citep{UGK98}. At the end of this phase, the NS has been certainly
heated up to surface temperatures $T_s\gtrsim 10^6$ K and the surface magnetic field has been reduced by
about four orders of magnitude (see e.g. Fig. 3 in \cite{UGK98}).\\
\noindent Clearly, accretion will cause both an accelerated magnetic field decay in the crust by heating and
an advection of the currents, that maintain the subsurface field, into deeper regions toward the crust-core
interface. In these layers, the magnetic field retaining currents can flow (almost) without resistance until 
the end of the accretion phase. These currents, located around the core crust interface, can form a very strong ($\gtrsim 10^{15}$ G) 
toroidal field structure \citep{GV14}. Such a configuration may exist there since the birth of the NS and/or has been formed by advection 
during the the Roche lobe overflow phase that spins up the MSP. 
Whether this field can significantly contribute to the heat budget of MSPs depends 
on the process of its re-diffusion towards the surface when accretion has been ceased. 
The corresponding processes have been described in detail e.g. by \cite{SG24}.
Recently, \cite{IEP16} studied in detail the process of re-diffusion of different magnetic field components after
the fall-back accretion on a newborn NS. They found that modes with $l >6$ re-diffuse after about $10^6$ yrs even 
to higher surface values than the dipolar or quadrupolar component, while the $l=1\ldots 6$ components retain
about one tenth of their original strength after $10^6$ years.\\
\noindent Of course, fall-back accretion is not the same as accretion during the Roche lobe overflow phase in a LMXB.
However, the length of the re-diffusion period is determined only by the depth of field submergence at the end of the
accretion phase and by the conductive properties of the crust \citep{GPZ99}. Hence, the re-diffusion will proceed similar 
in both cases and is certainly finished at J0437-4715.\\
A MSP that shows significant anisotropies in its surface temperature is J0030+0451 \citep{RWBRLGABPCGHHLMS19,
MLDBAGGHHLLMMRSWEFOPS19}. These authors report the certain evidence for the presence of higher order magnetic multipoles. 
Two clearly distinct surface temperatures are observed, a smaller area with $T_s\sim 2.7\times 10^6$ K
and a larger one with $T_s\sim 6.7\times 10^5$ K. Both temperatures are above those expected from cooling scenarios of normal NSs.
The hotter area certainly is the polar cap surface where strong and small scale field components must be necessarily persent.
Otherwise no radio emission can be produced\citep{GMG03}. For the larger area the heating mechanism discussed above may work.\\

\section{Conclusions}\label{sec4}
Today, the presence of small-scale magnetic field structures on the surface and in the upper layers of the crust of NSs 
is undisputed. Numerous observational results and theoretical studies confirm this. The idea that at the surface of MSPs 
the dominant magnetic field component is not the dipolar but a higher one with $l\ge 3$ has been discussed by \cite{K91}
in the context of MSP pulse morphology and field strengths observed in $\gamma$- ray bursts. The higher order
multipoles will not contribute much to the spin-down of the NSs but can affect their cooling history.\\
\noindent We have shown that, based on the formula of \cite{K91}, magnetic field components with multipolarity $l \ge 3$ cannot
be excluded as being sufficiently strong to heat MSPs to the observed surface temperatures of $T_s \gtrsim 10^5$ K.\\
The fact that no crustal fractures were observed on MSP sets an upper limit for the field strength of the multipoles. \cite{WH15} provided
a criterion that the magnetic pressure does not exceed that of the degenerated electrons. According to this criterion and given
a field strength of the multipoles of $\sim 10^{14}$ G, the crust may withstand the magnetic pressure as long as components
with $l=3$ or $l=4$ are present up to a density of $10^{10}$ g cm$^{-3}$, i.e. they should not be present above the bottom of the envelope.\\
\noindent This short study can not replace detailed numerical modelling of the small scale field structures and their influence
on the MSP cooling. However, it should be considered as an encouragement to perform such explorations.

\section*{Acknowledgments}
I thank to an anonymous referee for constructive criticism and valuable advices.\\
This work was supported by the grant 2020/37/B/ST9/02215 of the National 
Science Centre, Poland.

\bibliography{pulsars}

\end{document}